\documentclass{aastex6}

\shorttitle{The orbital ephemeris of the classical nova RR Pictoris}
\shortauthors{Vogt et al.}

\begin{document}


\title{The orbital ephemeris of the classical nova RR Pictoris: presence of a third body?}


\author{N. Vogt\altaffilmark{1}, M. R. Schreiber\altaffilmark{1}, F.-J. Hambsch\altaffilmark{3,4}, G. Retamales\altaffilmark{1}, C. Tappert\altaffilmark{1}, L. Schmidtobreick\altaffilmark{2} \& \\ I. Fuentes-Morales\altaffilmark{1}}

\affil{$^{1}$Instituto de F\'isica y Astronom\'ia, Universidad de Valpara\'iso, Valpara\'iso, Chile\\
$^{2}$European Southern Observatory, Santiago 19, Chile, Casilla 1900\\$^{3}$Vereniging Voor Sterrenkunde (VVS), Oude Bleken 12, 2400 Mol, Belgium \\$^{4}$ American Association of Variable Star Observers, 49 Bay State Rd., Cambridge, MA02138, USA }


\begin{abstract}
\noindent
{The ex-nova RR Pic presents a periodic hump in its light curve which is considered to refer to its orbital period. Analyzing all available epochs of these hump maxima in the literature, and combining them with those from new light curves obtained in 2013 and 2014, we establish an unique cycle count scheme valid during the past 50 years, and derive an ephemeris with the orbital period 0.145025959(15) days. The O - C diagram of this linear ephemeris reveals systematic deviations which could have different causes. One of them could be a light-travel-time effect caused by the presence of a hypothetical third body near the star/brown dwarf mass limit, with an orbital period of the order of 70 years.  We also examine the difficulty of the problematic of detecting sub-stellar or planetary companions of close red-dwarf white-dwarf binaries (including cataclysmic variables),  and discuss other possible mechanisms responsible for the observed deviations in O - C. For RR Pic, we propose strategies in order to solve this question by new observations.}
\end{abstract}


\section{Introduction}

\noindent
Classical Novae comprise an important subclass of cataclysmic variables (CVs), characterized by a single observed outburst of large amplitude (typically between 9 a 15 mag) which is understood as a thermonuclear runaway explosion on the surface of the white dwarf component. RR Pic is one of the brightest representatives of this class, erupting in 1925 and leaving a stellar remnant of about magnitude V $\approx$ 12 mag. Since $\sim$1960 several observers noticed a modulation in the light curve of this remnant, with a period near 3.5 hours, which was confirmed by \citet{vogt75} who presented photometric observations and derived an ephemeris of this periodic ``hump'', based on all data available at that time. Later, this ephemeris was improved by \citet{kubiak84} based on additional data. Time-resolved spectroscopic observations were published by \citet{wyckoff77} and by \citet{haefner91}; their radial velocity curves based on the dominant He {\sc ii} ($\lambda$ 4686\AA) emission line confirm that the photometric hump refers to the orbital period. Additional tomographic studies of the spectral behaviour were published by \citet{linda2003} and \citet{ribeiro2006}. 
\noindent
\citet{linda2008} presented new time-resolved photometric data, confirming the presence of the orbital hump and detecting, in addition, a superhump with a period excess of about 8.6\% over the orbital period.\\
Due to the large time gaps in the available observations previous authors were not able to determine a unique long-term ephemeris of the orbital period. Now, the main gap (between 1982 and 2005) could partly be filled by unpublished photometric observations in the CBA archive (kindly provided by J. Patterson and collaborators from the Centre for Backyard Astrophysics). In addition, we analyze here recent new observations obtained by one of us (F.-J.H.) in 2013 and 2014. The aim of our present paper is to derive, for the first time,  a unique ephemeris for the orbital hump of RR Pic valid during the last five decades (section \ref{s2}). In this context some systematic deviations in the O - C diagram were detected which are discussed tentatively as light-travel-time effect in Section \ref{s3}. In Sections \ref{s4} and \ref{s5} we compare the third-body hypothesis of RR Pic with the actual situation of other known or suspected multiple stellar systems (including binary star-planet configurations), mention possible alternative interpretations and add some conclusions.

\section{ New observations and determination of a long-term orbital ephemeris } 
\label{s2}
\begin{figure}
  \centering  
  \resizebox{0.7\textwidth}{!}{\includegraphics{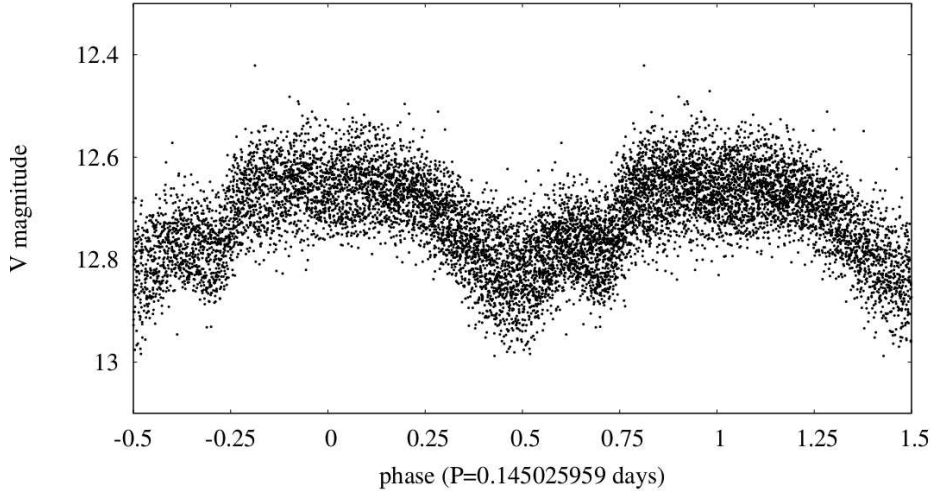}}
  \caption{All ROAD observations from 2013 and 2014 vs. phase of the ephemeris (\ref{eq1}). \label{fig1}}
\end{figure}

\noindent
It was essential to obtain a significant amount of new time-resolved photometric data of RR Pic, in order to achieve a unique way to count the cycles of the orbital humps. These observations were carried out at the Remote Observatory Atacama Desert (ROAD), located in San Pedro de Atacama, Chile. This observatory contains a 40-cm telescope (f/6.8) from Orion optics, England, and a camera with a Kodak 16803 CCD chip with 4k$\times$4k pixels of 9 $\mu$m size provided by Finger Lakes Instrumentation (FLI) \citep{hambshroad2012}, operating in a robotic mode. RR Pic was observed in the V band during a total of 36 nights, between 2013, February, 27 and April 1, and in 81 nights between 2013, Nov. 21 and 2014, March 15, with a typical time resolution between 2 and 3 minutes. Fig. \ref{fig1} shows the mean light curve vs. phase determined from all ROAD observations in 2013 and 2014. Similarities and differences of its shape, compared to mean light curves at earlier epochs, is being discussed by Fuentes et al. 2016 in preparation.

The method to determine the individual epoch of each hump maximum in our new light curves was not very sophisticated: Since a typical hump amplitude is between 0.2 and 0.3 mag we considered, in most cases, as the hump maximum epoch the average time between the ascending and the descending branch of the hump light curve, at a level about 0.1 mag below the maximal brightness. Only if a hump appeared rather asymmetric, with a pronounced real maximal magnitude occurring before or after the mentioned symmetric hump, we took the average between ``real'' and ``symmetric'' time as the hump maximum epoch. The same method for determining hump maximum timings was used for light curves in publications (or unpublished data sets) which only are available in figures or as digital data.  If authors had determined and published hump maximum times, we used their values.  It should be emphasized that the the method of how to determine individual hump epochs does not influence significantly the long-term periodic behaviour if any uncertainty in correctly counting the cycles can be avoided. The observed hump often is superimposed by strong flickering which causes an unavoidable scatter in the O - C diagram, typically between 0.005 and 0.008 days in the RR Pic data. The possible presence of superhumps does not affect the overall ephemeris because their amplitude is about a factor 10 smaller than that of the orbital hump (\citealt{linda2008}; Fuentes et al. 2016 in preparation.

We have subdivided the data into 8 sets: Set A - C refer to \citet{kubiak84} and references therein. Set B and C contains also additional maximum epochs determined by us from figs. 1, 2, 3 and 4 in \citet{warner86}. These three sets comprise the observations obtained between 1965 and 1982. Sets D, E and G  refer to hump maxima obtained from unpublished observations by Joe Patterson and collaborators at different observatories, obtained between 1999 and 2007 and collected in the data base of CBA (Centre for Backyard Astrophysics). The hump maxima in Set E have been derived from fig. 1 of \citet{linda2008}, obtained in 2005, while set H contains our recent ROAD observations from 2013/14. 

In spite of the large gaps between subsequent sets of observations, it was possible to determine a unique cycle counting system. We  performed linear least squares fits HJD vs E in several steps, first within each set, then combining adjacent sets of hump maxima times, and derived a period value for sets A-C of 0.14502533(7) days and for D-H of 0.14502609(4) days; finally, these both period values were sufficient accurate to bridge the large gap between sets C and D in  a unique way. This procedure led to the linear ephemeris 
\begin{equation}
\textup{HJD(max.)}=2438815.3664(15) + 0.145025959(15)\cdot\mathrm{E}
\label{eq1}
\end{equation}  
\noindent
with a standard deviation of 0.0079 days, based on a total of 203 hump maxima of RR Pic. Fig. \ref{fig2} shows the mean O - C values for each of the data sets A -- H. Table \ref{tbl-1} lists the total ranges in E, the mean O - C values and their errors together with the corresponding references. A parabolic least-squares fit over the same 203 hump maximum epochs reveals
\begin{equation}
\textup{HJD(max.)}  =  2438815.3769(22) + 0.145025500(73)\cdot\textup{E}  + (32\pm5)\cdot10^{-13} \cdot\textup{E}^{2}  
\label{eq2}
\end{equation} 

\noindent
(standard deviation 0.0072 days). A table with all 203 epochs E and their HJD values of hump maxima is given  the on-line data annex. 

The parabolic term in equation \ref{eq2}, which is significant at $\sim$6$\sigma$ level, indicates a changing period. However, this fact crucially depends on the negative mean O - C value of set C from the linear ephemeris \ref{eq1}. Set C contains a total of 19 hump maximum observations from three different sources: 9 of them are maximum timing values published by \citet{haefner82} and other 4 values were published by \citet{kubiak84}; the remaining 6 maximum timings have been determined by us from the light curves of \citet{warner86}. Therefore, there are three independent sources coinciding in rather strongly negative O - C values at the epoch of set C, reinforcing the validity of the period change reported here.

The epochs listed in Table \ref{tbl-1} cover in total  more than 49 years; the initial epoch coincides with the end of the recovery from the nova outburst. We have extracted all visual observations published in the AAVSO archive for these five decades. A linear least squares fit through these data reveals that RR Pic was declining in visual brightness at a mean rate of (1.89 $\pm$ 0.07)$\cdot10^{-5}$  mag/day or about 0.34 mag in the 50-year time interval. This can be compared to other classical novae: \citet{johnson2014} determined a decline rate of 0.44 $\pm$ 0.04 mag per century for Nova Aquilae 1918 (V 603 Aql).

\begin{deluxetable}{lrrrrcr}

\tablewidth{0pt}
\tablecaption{Mean $\overline{\rm (O-C)}$ deviations of the hump maxima in each of the 8 sets of observations at different epochs, based on ephemeris (\ref{eq1}). Each set is characterized by its first (E$_{\rm start}$) and last (E$_{\rm end}$) hump maximum epoch. N is the number of single hump observations in each set, $\sigma\overline{\rm (O-C)}$ the mean errors of the mean O - C deviations. \label{tbl-1}}
\tablehead{
\colhead{Set}  & \colhead{N}  &  \colhead{E$_{\rm start}$}           & \colhead{E$_{\rm end}$}  &
\colhead{$\overline{\rm (O-C)}$}     &  \colhead{$\sigma\overline{\rm (O-C)}$}    & \colhead{Ref.}\\
               &              &                                      &                          &            10$^{-4}$d           &  10$^{-4}$d                  &       }
{\startdata{
A  &      7    &       0   &   10041   &   +105.0  &     9.6  &     1    \\
B  &      16   &   19590   &   26969   &   +59.8   &    17.4  &     1, 2 \\
C  &      19   &   37513   &   44395   &   -95.8   &    14.7  &     1, 2 \\
D  &      19   &   85479   &   90685   &   -38.1   &    17.2  &     4a   \\
E  &      20   &   100394  &   100552  &   +1.3    &    10.6  &     4b   \\
F  &      12   &   100632  &   101052  &   -37.3   &    11.0  &     3    \\
G  &       8   &   105490  &   105634  &   +6.3    &    14.5  &     4c   \\
H  &     102   &   120862  &   123531  &   +12.3   &     7.3  &     5    \\
}\enddata}

\tablerefs{
(1) \citealt{kubiak84} and references therein ; (2) \citealt{warner86}; (3) \citealt{linda2008};
(4a) CBA, priv. commun. Set1; (4b) CBA, priv. commun. Set2 (4c)  CBA, priv. commun. Set3,
(5) new ROAD observations}

\end{deluxetable}

\section{Possible interpretation as light-travel-time effect and properties of a hypothetical third body}
\label{s3}
\begin{figure}
  \centering  
  \resizebox{0.7\textwidth}{!}{\includegraphics[trim={0mm 60mm 00mm 60mm}, clip]{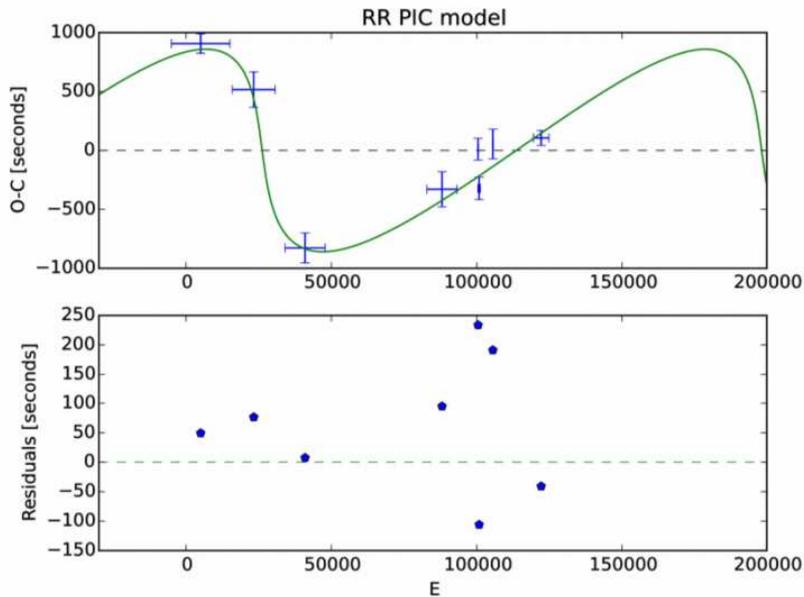}}
  \caption{Upper panel: The averaged O - C deviations vs. mean E of the hump maximum epochs according to ephemeris (1) for each data set A -- H according to Table \ref{tbl-1}. The vertical bars refer to the errors of the mean O - C values while the horizontal bars mark the ranges in E for each data set. The green curve refers to the third-body model described in the text and in Table \ref{tbl-2}. Lower panel: Residuals from the model O - C curve. \label{fig2}}
\end{figure}

\noindent
In Fig.\ref{fig2} (upper panel) we present an O - C diagram based on ephemeris \ref{eq1}, corresponding to the mean values listed in Table \ref{tbl-1}. Apparently, the orbital period of RR Pic during the last five decades was not constant. In the first 18 years (sets A, B and C, E $<$ 50000) the period value was smaller than the overall average, and after the gap without any available observations the period was larger in set D -- H (E $>$ 85000). This general trend is confirmed by the parabolic fit \ref{eq2} which reveals a tendency of an increasing period. One of the possible interpretations would be the presence of a third body, in an eccentric orbit, causing a light-travel-time effect on the periodic hump behaviour of the ex-nova RR Pic. We are aware that the presently available data do not prove this hypothesis, but we would like to present here a first estimation of possible parameters of such a hypothetical companion, and give predictions for future observations, in order to prove or to refute this hypothesis.

For a rough estimation of possible parameters of the third body we use the Monte-Carlo Markow Chain (MCMC) code described in more detail in \citet{hardy2015} for deriving possible orbital parameters of a third body around an eclipsing binary system from O - C observations. We fitted the O - C distribution listed in Table \ref{tbl-1} and used a fixed value for sum of the masses of the primary and secondary components of the CV RR Pic, i.e. M$_1$ + M$_2$ = 1.35 solar masses, based on M$_1$ = 0.95 M$_{\odot}$ and M$_2$ = 0.4 M$_{\odot}$, according to \citet{haefner82}. First, we determined  appropriate values of the argument of periastron $\omega$ and the orbital inclination \textit{i}. Then we performed calculations with different trial values of the eccentricity \textit{e} and found a minimum in the residuals. The corresponding model parameters are given in Table \ref{tbl-2}. The model is presented in Fig. \ref{fig2} as a line, and the residuals are also shown in the lower panel of Fig. \ref{fig2}.

\begin{table}
\small
\centering
\caption{Model parameters of the hypothetic third body in RR Pic. \label{tbl-2}}
\scalebox{0.83} {
\begin{tabular}{lr}
\tableline\tableline
Orbital period around the CV RR Pic   &  P$_3$  =  66.7  $\pm$ 2.5   yrs          \\
Third body mass                       &  M$_3$  =  0.258 $\pm$ 0.023  M$_{\odot}$ \\ 
Mean distance to the CV RR Pic        &  a      = 19.25  $\pm$ 0.46 AU            \\
Orbital inclination                   &  {\it i} = 90$^{\circ}$   adopted           \\
Orbital eccentricity                  &  {\it e} = 0.8357  adopted                 \\
Argument of periastron                & $\omega$= 179.7$^{\circ}$   adopted       \\
Julian date of the periastron passage &   T$_p$ = 2442827 $\pm$  283 d            \\
\tableline \tableline
\end{tabular}
}

\end{table}

Apparently, the observations presented here do not cover an entire orbital revolution of the third body;  its mass corresponds to a mid-M dwarf. The error ranges listed in Table \ref{tbl-2} do not include the uncertainty in the mass of the CV RR Pic.

The distance of RR Pic has been determined applying astrometry with the HST by \citet{harrison2013}, revealing a parallax of 1.92 $\pm$ 0.18 mas which corresponds to 521 $\pm$ 50 pc. At this distance the maximal projected mean angular distance between the CV RR Pic and the third body would be of the order of 40 mas, just below the limit for a direct detection via imaging. For instance, the low mass stellar companions of young members of the Cha star-forming regions could be detected with the VLT and NACO at projected distances down to 50 mas, depending also on the magnitude difference between host star and companion \citep{vogt2012}. In RR Pic this difference is probably $\geq$ 3 mag in the K band. This is just on the limit of the Sparse Aperture Masking (SAM) interferometric technology applied at NACO/VLT in the Paranal observatory \citep{lacour2011}. On the other hand, since the hypothetical third body is moving on a rather eccentric orbit its projected angular distance could vary strongly and increase its value at certain phases, facilitating a direct imaging detection. 

An astrometric detection of the movement of the CV RR Pic around the center of mass of the triple system should be possible, using similar techniques as those of \citet{harrison2013}, after observing the field for decades. From the large mass ratio between the CV and the third body, the $\gamma$ radial velocity of the CV RR Pic is expected to vary in the range $\leq$ 2 km/s at time scales of decades, probably difficult to detect due to the natural variability in the CV spectrum, especially its accretion disc. On the other hand, at a mass ratio of M$_1$/M$_2$ $\approx$ 2.4, for the CV RR Pic, and based on its  radial velocity semi amplitude 120 km/s $\leq$ K$_1$ $\leq$ 170 km/s (\citealt{haefner91}; \citealt{linda2003}) we expect K$_2$ $\approx$ 350 km/s for the secondary star in RR Pic. If the secondary star and the third body would be of similar luminosity at infrared wave lengths and if their spectrum could be observed, we would expect a superposition of two late type star lines or bands, one of them stationary, the other one showing rather strong Doppler shifts with the orbital phase of the CV RR Pic. However, K-band spectroscopy by \citet{harrison2005} found that the accretion disc still contributes about 80\% to the flux even in that wavelength range. While their data suggest the presence of CO absorption features from the secondary star, they appear too weak to be of much use for this purpose even using high-resolution spectroscopy.

\section{Comparison to PCEB\MakeLowercase{s} and discussion of alternatives interpretations}
\label{s4}
\noindent
During the last years eclipse timing variations have been identified in nine eclipsing post-common-envelope binary stars (PCEBs) and three polars (AM Her type CVs) \citep{zorotovic2013}. The most frequently published interpretation for this phenomenon is the presence of one or two circumbinary sub-stellar objects orbiting the host compact binary system. In this section we review possible light-travel-time effects in PCEBs, compared to other explanations of period variations, and apply them to our new results on RR Pic.

Support for the third body interpretation of apparent period changes comes mostly from the detached PCEB  NN Ser. \citet{marsh2014} and \citet{Beuermann2013} have shown that new high precision eclipse timing measurements are still in agreement with the two-planet model for NN Ser proposed by \citet{beuermann2010}. In addition, alternative explanations can be excluded for NN Ser. As shown by \citet{Brinkworth2006}, the M-dwarf companion in NN Ser does not provide the required energy to drive the period variations via Applegate's mechanism, i.e. modulations by the gravitational coupling of the orbit to variations in the shape of a magnetically active star in the system \citep{applegate92}. In addition, the timing data for the secondary eclipses in NN Ser follow the same trend seen in the primary eclipse times, ruling out apsidal precession as a possible cause for the variations \citep{parsons2014}.

However, NN Ser remains an exception. In other PCEBs with timing variations the situation is quite different, because  there are often serious problems with the third body model. First, some suggested planetary systems turned out to be dynamically unstable \citep{hinse2012} and others drastically disagreed with more recent high precision eclipse timings \citep{Parsons2010}. Second, \citet{hardy2015} observed the eclipsing PCEB V471 Tau with SPHERE aiming for a detection of the claimed circumbinary brown dwarf but did not detect it despite clearly reaching the required contrast. Thus, despite the quite convincing case NN Ser, it might well be that an alternative and so far unknown mechanism is driving the eclipse timing variations in PCEBs and perhaps also in RR Pic.

In the case of the planets around PCEBs, if they exist at all, the origin of the orbiting third bodies, remains unclear. The hypothetical third objects must have either survived the common envelope (CE) evolution of the host binary star or they must have formed following the common envelope phase in a second generation scenario \citep{perets2010}. Both scenarios are far from straightforward. If the third objects formed with the binary stars, it remains to be explained why virtually all PCEBs with good timing coverage show variations \citep{zorotovic2013} while only 1-3 per cent of main sequence binaries host giant planets \citep{Welsh2012}. If, on the other hand, the third objects formed from the mass ejected during common envelope evolution, the formation process needs to be very fast as some of the primary white dwarfs have cooling ages less than a million year (e.g. NN Ser). This is very short compared to the $\gtrsim$ 4 Myr predicted by the core accretion model for giant planet formation and would require gravitational instabilities in a second generation disk as proposed by \citet{schleicher2014}. This second generation planet formation idea is supported by the recent finding of \citet{hardy2016} who detected cold dust around NN Ser with ALMA and concluded that it is most likely material left over from common envelope evolution.

In summary, the third body model does so far not provide a convincing general explanation for the observed period variations in compact binary stars. It’s main strengths are the fact that is seems to work well for NN Ser and that we do not have an alternative explanation. However, as stated  in \citet{marsh2014}, the latter may well mean that we have just not been clever enough yet.

The situation for RR Pic is different to the eclipse timing variations previously reported in PCEBs and CVs because we find that a stellar third object is required to drive the large amplitude variations we see in RR Pic. Therefore, it is clear that the third object cannot have formed from material remaining bound to the central binary following CE evolution. Instead, the third star must have formed together with the central binary and survived the CE evolution of the central objects. As hierarchical triple systems represent a rather frequent configuration of main sequence stars, this scenario appears to be reasonable for RR Pic. In fact, given that \citep{tokovinin2014a, tokovinin2014b} and \citet{Lohr2015} estimate the higher order multiplicity fraction among binaries to be approximately one quarter, CVs with stellar companions at wide separations should be expected to exist and RR Pic might be the first convincing candidate for such systems. This interpretation is also consistent with the large eccentricity required to fit the O - C diagram of RR Pic. As the mass loss during CE evolution is supposed to be very fast (e.g. \citealt{Webbink2008},   \citealt{zorotovic2010}), it can be considered as adiabatic and one would expect a significant increase in eccentricity for the orbit of the third body.

While the third body hypothesis appears a reasonable explanation for the
timing variations in RR Pic, existing alternative explanations for PCEBs
clearly fail to drive period variations as strong as those of RR Pic. The
energy required to generate period changes of about 1000 seconds via the
Applegate mechanism clearly exceeds the energy available in the secondary star
(see \citealt{Parsons2010} for details). Apsidal motion can be excluded as
well as the eccentricity required to explain timing variations of 1000 seconds
for the parameters of RR Pic is $\sim$ 0.1 (using equation 2 of
\citealt{barlow2012}). Such a high eccentricity can be clearly excluded for
old short orbital period systems such as CVs. A possible alternative
  scenario that might only apply to CVs is related to the nature of RR\,Pic as
  a post nova. In order to explain the white dwarf mass problem in CVs \citep{zorotovicetal2011},
  it has been suggested that classical nova eruptions may in some cases cause
  significant frictional angular momentum loss \citep[][]{schreiberetal2016,nelemansetal2016}. 
In principal, strong additional angular momentum loss following the nova eruption
could explain the observed O - C diagram. The steep decline the O - C shows
during the first $\sim\,15$ years of data could be part of a parabolic term that 
has been caused by strong angular momentum loss. The fact that the shell has still been 
detected 70 years after the eruption and the expansion velocity was rather
slow \citep{gill+obrian1998} 
may support this idea because slower expansion velocities are expected to
generate stronger frictional angular momentum loss 
\citep{schenkeretal98}. However, given that our oldest data has been taken
$\sim\,40$ years after the nova eruption, the possible connection between the
nova eruption and the observed O - C variations remains speculative. 

\section{Conclusions: third bodies in CVs?}
\label{s5}
\noindent
Based on a rather stable periodic orbital hump in the light curve of RR Pic
between 1965 and 2014, we present a unique and precise ephemeris, valid for
these five decades. In addition, we detected rather strong orbital period
variations. In particular, there is a significant difference between the mean
period valid in 1965--1982, and that afterwards, in the sense of an increasing
period value during the total time interval of five decades covered. As one of
the possible interpretations of these period variations we propose a
light-travel-time effect and estimate that it could be caused by a hypothetic
late M type companion with  M$_3$ = 0.26 $\pm$ 0.02  solar masses, completing
an eccentric orbit around the CV RR Pic in $\approx$ 67 years at a mean
distance from the CV of  $\approx$19 AU. 

Among the CVs, only for the three eclipsing polars planetary candidates have been reported:  UZ For  \citep{potter2011}, HU Aqr \citep{schwope2011} and DP Leo \citep{beuermann2011}. Reports on triple systems in other CVs, also involving stars and/or brown dwarf components, are very rare. To our knowledge, there are only two candidates for such triple systems with  CV components: the first
case is FS Aur, a dwarf nova and intermediate polar \citep{Neustroev2013}. For
this star, a long-term modulation in its light curve with a period of
$\approx$ 900 days was detected and interpreted by means of the interaction
with a third body of a mass between 25 and 64 times that of Jupiter
\citep{chavez2012}. All these four cases have ultra-short orbital periods
($<$0.088 d). The second case of a possible triple system is FH Leo, a wide
visual binary consisting on two F/G type main sequence stars, at an angular
separation of about 8 seconds of arc. HIPPARCOS had detected some possible
flares without resolving the binary. \citet{vogt2006} suggested that these
could be caused by a SU UMa type dwarf nova orbiting around one of the two
double star components. The latter two cases, which do not involve
light-travel-time effect, are rather hypothetic, without a stringent
confirmation.  

We would like to emphasize that our interpretation of the long-term O - C
diagram of RR Pic, if confirmed, would imply, for the first time, the
detection of a cataclysmic binary with a third stellar companion, and also the
first candidate for a triple system CV above the period gap between 2 and 3
hours. In addition, the case RR Pic would be the first application of the
light-travel-time method in a non-eclipsing binary system. This was only
possible because the amplitude of this effect, in case of a stellar companion,
is about one order of magnitude larger than for planetary companions. Finally,
RR Pic would be the first non-magnetic CV in a triple system. Therefore, we
believe that it could be important to present these results in spite of the
rather hypothetical character of our interpretation. It remains an open
question why just RR Pic seems to be a record holder in so many respects, in
spite of its rather normal behaviour as a non-magnetic CV. There are many
other CVs with orbital hump light curves, and well defined eclipses,
especially those with separate ingress and egress eclipse phases of the white
dwarf, as present in Z Cha, OY Car and HT Cas among others. Does none of them
possess a third body? Perhaps a re-analysis of the available data on eclipsing CVs might result in further discoveries.

\acknowledgments
We thank the CBA team, especially Joe Patterson and J. Kemp, for making unpublished observations of RR Pic 1999 - 2007 available to us. This research was supported by grant FONDECYT 1120338 (CT, NV) and DIUV 38/2011 (NV). MRS thanks for support from Fondecyt (1141269) and the and Millennium Nucleus RC130007 (Chilean Ministry of Economy). NV, MRS, CT and IFM acknowledge support by the Centro de Astrof\'isica de Valpara\'iso (CAV). 

\bibliographystyle{aasjournal}
\bibliography{RRPic_third_body}

\newcommand{\noop}[1]{}
\begin{thebibliography}{}
\expandafter\ifx\csname natexlab\endcsname\relax\def\natexlab#1{#1}\fi

\bibitem[{{Applegate}(1992)}]{applegate92}
{Applegate}, J.~H. 1992, \apj, 385, 621

\bibitem[{{Barlow} {et~al.}(2012){Barlow}, {Wade}, \& {Liss}}]{barlow2012}
{Barlow}, B.~N., {Wade}, R.~A., \& {Liss}, S.~E. 2012, \apj, 753, 101

\bibitem[{{Beuermann} {et~al.}(2013){Beuermann}, {Dreizler}, \&
  {Hessman}}]{Beuermann2013}
{Beuermann}, K., {Dreizler}, S., \& {Hessman}, F.~V. 2013, \aap, 555, A133

\bibitem[{{Beuermann} {et~al.}(2010){Beuermann}, {Hessman}, {Dreizler},
  {Marsh}, {Parsons}, {Winget}, {Miller}, {Schreiber}, {Kley}, {Dhillon},
  {Littlefair}, {Copperwheat}, \& {Hermes}}]{beuermann2010}
{Beuermann}, K., {Hessman}, F.~V., {Dreizler}, S., {et~al.} 2010, \aap, 521,
  L60

\bibitem[{{Beuermann} {et~al.}(2011){Beuermann}, {Buhlmann}, {Diese},
  {Dreizler}, {Hessman}, {Husser}, {Miller}, {Nickol}, {Pons}, {Ruhr},
  {Schm{\"u}lling}, {Schwope}, {Sorge}, {Ulrichs}, {Winget}, \&
  {Winget}}]{beuermann2011}
{Beuermann}, K., {Buhlmann}, J., {Diese}, J., {et~al.} 2011, \aap, 526, A53

\bibitem[{{Brinkworth} {et~al.}(2006){Brinkworth}, {Marsh}, {Dhillon}, \&
  {Knigge}}]{Brinkworth2006}
{Brinkworth}, C.~S., {Marsh}, T.~R., {Dhillon}, V.~S., \& {Knigge}, C. 2006,
  \mnras, 365, 287

\bibitem[{{Chavez} {et~al.}(2012){Chavez}, {Tovmassian}, {Aguilar}, {Zharikov},
  \& {Henden}}]{chavez2012}
{Chavez}, C.~E., {Tovmassian}, G., {Aguilar}, L.~A., {Zharikov}, S., \&
  {Henden}, A.~A. 2012, \aap, 538, A122

\bibitem[{{Gill} \& {O'Brien}(1998)}]{gill+obrian1998}
{Gill}, C.~D., \& {O'Brien}, T.~J. 1998, \mnras, 300, 221

\bibitem[{{Haefner} \& {Betzenbichler}(1991)}]{haefner91}
{Haefner}, R., \& {Betzenbichler}, W. 1991, Information Bulletin on Variable
  Stars, 3665, 1

\bibitem[{{Haefner} \& {Metz}(1982)}]{haefner82}
{Haefner}, R., \& {Metz}, K. 1982, \aap, 109, 171

\bibitem[{{Hambsch}(2012)}]{hambshroad2012}
{Hambsch}, F.-J. 2012, Journal of the American Association of Variable Star
  Observers (JAAVSO), 40, 1003

\bibitem[{{Hardy} {et~al.}(2015){Hardy}, {Schreiber}, {Parsons}, {Caceres},
  {Retamales}, {Wahhaj}, {Mawet}, {Canovas}, {Cieza}, {Marsh}, {Bours},
  {Dhillon}, \& {Bayo}}]{hardy2015}
{Hardy}, A., {Schreiber}, M.~R., {Parsons}, S.~G., {et~al.} 2015, \apjl, 800,
  L24

\bibitem[{{Hardy} {et~al.}(2016){Hardy}, {Schreiber}, {Parsons}, {Caceres},
  {Brinkworth}, {Veras}, {G{\"a}nsicke}, {Marsh}, \& {Cieza}}]{hardy2016}
---. 2016, \mnras, 459, 4518

\bibitem[{{Harrison} {et~al.}(2013){Harrison}, {Bornak}, {McArthur}, \&
  {Benedict}}]{harrison2013}
{Harrison}, T.~E., {Bornak}, J., {McArthur}, B.~E., \& {Benedict}, G.~F. 2013,
  \apj, 767, 7

\bibitem[{{Harrison} {et~al.}(2005){Harrison}, {Osborne}, \&
  {Howell}}]{harrison2005}
{Harrison}, T.~E., {Osborne}, H.~L., \& {Howell}, S.~B. 2005, \aj, 129, 2400

\bibitem[{{Hinse} {et~al.}(2012){Hinse}, {Lee}, {Go{\'z}dziewski},
  {Haghighipour}, {Lee}, \& {Scullion}}]{hinse2012}
{Hinse}, T.~C., {Lee}, J.~W., {Go{\'z}dziewski}, K., {et~al.} 2012, \mnras,
  420, 3609

\bibitem[{{Johnson} {et~al.}(2014){Johnson}, {Schaefer}, {Kroll}, \&
  {Henden}}]{johnson2014}
{Johnson}, C.~B., {Schaefer}, B.~E., {Kroll}, P., \& {Henden}, A.~A. 2014,
  \apjl, 780, L25

\bibitem[{{Kubiak}(1984)}]{kubiak84}
{Kubiak}, M. 1984, \actaa, 34, 331

\bibitem[{{Lacour} {et~al.}(2011){Lacour}, {Tuthill}, {Ireland}, {Amico}, \&
  {Girard}}]{lacour2011}
{Lacour}, S., {Tuthill}, P., {Ireland}, M., {Amico}, P., \& {Girard}, J. 2011,
  The Messenger, 146, 18

\bibitem[{{Lohr} {et~al.}(2015){Lohr}, {Norton}, {Payne}, {West}, \&
  {Wheatley}}]{Lohr2015}
{Lohr}, M.~E., {Norton}, A.~J., {Payne}, S.~G., {West}, R.~G., \& {Wheatley},
  P.~J. 2015, \aap, 578, A136

\bibitem[{{Marsh} {et~al.}(2014){Marsh}, {Parsons}, {Bours}, {Littlefair},
  {Copperwheat}, {Dhillon}, {Breedt}, {Caceres}, \& {Schreiber}}]{marsh2014}
{Marsh}, T.~R., {Parsons}, S.~G., {Bours}, M.~C.~P., {et~al.} 2014, \mnras,
  437, 475

\bibitem[{{Nelemans} {et~al.}(2016){Nelemans}, {Siess}, {Repetto}, {Toonen}, \&
  {Phinney}}]{nelemansetal2016}
{Nelemans}, G., {Siess}, L., {Repetto}, S., {Toonen}, S., \& {Phinney}, E.~S.
  2016, \apj, 817, 69

\bibitem[{{Neustroev} {et~al.}(2013){Neustroev}, {Tovmassian}, {Zharikov}, \&
  {Sjoberg}}]{Neustroev2013}
{Neustroev}, V.~V., {Tovmassian}, G.~H., {Zharikov}, S.~V., \& {Sjoberg}, G.
  2013, \mnras, 432, 2596

\bibitem[{{Parsons} {et~al.}(2010){Parsons}, {Marsh}, {Copperwheat}, {Dhillon},
  {Littlefair}, {Hickman}, {Maxted}, {G{\"a}nsicke}, {Unda-Sanzana}, {Colque},
  {Barraza}, {S{\'a}nchez}, \& {Monard}}]{Parsons2010}
{Parsons}, S.~G., {Marsh}, T.~R., {Copperwheat}, C.~M., {et~al.} 2010, \mnras,
  407, 2362

\bibitem[{{Parsons} {et~al.}(2014){Parsons}, {Marsh}, {Bours}, {Littlefair},
  {Copperwheat}, {Dhillon}, {Breedt}, {Caceres}, \& {Schreiber}}]{parsons2014}
{Parsons}, S.~G., {Marsh}, T.~R., {Bours}, M.~C.~P., {et~al.} 2014, \mnras,
  438, L91

\bibitem[{{Perets}(2010)}]{perets2010}
{Perets}, H.~B. 2010, ArXiv e-prints, arXiv:1001.0581

\bibitem[{{Potter} {et~al.}(2011){Potter}, {Romero-Colmenero}, {Ramsay},
  {Crawford}, {Gulbis}, {Barway}, {Zietsman}, {Kotze}, {Buckley}, {O'Donoghue},
  {Siegmund}, {McPhate}, {Welsh}, \& {Vallerga}}]{potter2011}
{Potter}, S.~B., {Romero-Colmenero}, E., {Ramsay}, G., {et~al.} 2011, \mnras,
  416, 2202

\bibitem[{{Ribeiro} \& {Diaz}(2006)}]{ribeiro2006}
{Ribeiro}, F.~M.~A., \& {Diaz}, M.~P. 2006, \pasp, 118, 84

\bibitem[{{Schenker} {et~al.}(1998){Schenker}, {Kolb}, \&
  {Ritter}}]{schenkeretal98}
{Schenker}, K., {Kolb}, U., \& {Ritter}, H. 1998, \mnras, 297, 633

\bibitem[{{Schleicher} \& {Dreizler}(2014)}]{schleicher2014}
{Schleicher}, D.~R.~G., \& {Dreizler}, S. 2014, \aap, 563, A61

\bibitem[{{Schmidtobreick} {et~al.}(2008){Schmidtobreick}, {Papadaki},
  {Tappert}, \& {Ederoclite}}]{linda2008}
{Schmidtobreick}, L., {Papadaki}, C., {Tappert}, C., \& {Ederoclite}, A. 2008,
  \mnras, 389, 1345

\bibitem[{{Schmidtobreick} {et~al.}(2003){Schmidtobreick}, {Tappert}, \&
  {Saviane}}]{linda2003}
{Schmidtobreick}, L., {Tappert}, C., \& {Saviane}, I. 2003, \mnras, 342, 145

\bibitem[{{Schreiber} {et~al.}(2016){Schreiber}, {Zorotovic}, \&
  {Wijnen}}]{schreiberetal2016}
{Schreiber}, M.~R., {Zorotovic}, M., \& {Wijnen}, T.~P.~G. 2016, \mnras, 455,
  L16

\bibitem[{{Schwope} {et~al.}(2011){Schwope}, {Horne}, {Steeghs}, \&
  {Still}}]{schwope2011}
{Schwope}, A.~D., {Horne}, K., {Steeghs}, D., \& {Still}, M. 2011, \aap, 531,
  A34

\bibitem[{{Tokovinin}(2014{\natexlab{a}})}]{tokovinin2014a}
{Tokovinin}, A. 2014{\natexlab{a}}, \aj, 147, 86

\bibitem[{{Tokovinin}(2014{\natexlab{b}})}]{tokovinin2014b}
---. 2014{\natexlab{b}}, \aj, 147, 87

\bibitem[{{Vogt}(1975)}]{vogt75}
{Vogt}, N. 1975, \aap, 41, 15

\bibitem[{{Vogt}(2006)}]{vogt2006}
---. 2006, \aap, 452, 985

\bibitem[{{Vogt} {et~al.}(2012){Vogt}, {Schmidt}, {Neuh{\"a}user}, {Bedalov},
  {Roell}, {Seifahrt}, \& {Mugrauer}}]{vogt2012}
{Vogt}, N., {Schmidt}, T.~O.~B., {Neuh{\"a}user}, R., {et~al.} 2012, \aap, 546,
  A63

\bibitem[{{Warner}(1986)}]{warner86}
{Warner}, B. 1986, \mnras, 219, 751

\bibitem[{{Webbink}(2008)}]{Webbink2008}
{Webbink}, R.~F. 2008, in Astrophysics and Space Science Library, Vol. 352,
  Astrophysics and Space Science Library, ed. E.~F. {Milone}, D.~A. {Leahy}, \&
  D.~W. {Hobill}, 233

\bibitem[{{Welsh} {et~al.}(2012){Welsh}, {Orosz}, {Carter}, {Fabrycky}, {Ford},
  {Lissauer}, {Pr{\v s}a}, {Quinn}, {Ragozzine}, {Short}, {Torres}, {Winn},
  {Doyle}, {Barclay}, {Batalha}, {Bloemen}, {Brugamyer}, {Buchhave},
  {Caldwell}, {Caldwell}, {Christiansen}, {Ciardi}, {Cochran}, {Endl},
  {Fortney}, {Gautier}, {Gilliland}, {Haas}, {Hall}, {Holman}, {Howard},
  {Howell}, {Isaacson}, {Jenkins}, {Klaus}, {Latham}, {Li}, {Marcy}, {Mazeh},
  {Quintana}, {Robertson}, {Shporer}, {Steffen}, {Windmiller}, {Koch}, \&
  {Borucki}}]{Welsh2012}
{Welsh}, W.~F., {Orosz}, J.~A., {Carter}, J.~A., {et~al.} 2012, \nat, 481, 475

\bibitem[{{Wyckoff} \& {Wehinger}(1977)}]{wyckoff77}
{Wyckoff}, S., \& {Wehinger}, P.~A. 1977, in IAU Colloq. 42: The Interaction of
  Variable Stars with their Environment, ed. R.~{Kippenhahn}, J.~{Rahe}, \&
  W.~{Strohmeier}, 201

\bibitem[{{Zorotovic} \& {Schreiber}(2013)}]{zorotovic2013}
{Zorotovic}, M., \& {Schreiber}, M.~R. 2013, \aap, 549, A95

\bibitem[{{Zorotovic} {et~al.}(2011){Zorotovic}, {Schreiber}, \&
  {G{\"a}nsicke}}]{zorotovicetal2011}
{Zorotovic}, M., {Schreiber}, M.~R., \& {G{\"a}nsicke}, B.~T. 2011, \aap, 536,
  A42

\bibitem[{{Zorotovic} {et~al.}(2010){Zorotovic}, {Schreiber}, {G{\"a}nsicke},
  \& {Nebot G{\'o}mez-Mor{\'a}n}}]{zorotovic2010}
{Zorotovic}, M., {Schreiber}, M.~R., {G{\"a}nsicke}, B.~T., \& {Nebot
  G{\'o}mez-Mor{\'a}n}, A. 2010, \aap, 520, A86

\end{thebibliography}





\end{document}